\documentclass[twocolumn,showpacs,pre,aps,superscriptaddress]{revtex4}
\usepackage[utf8]{inputenc}
\usepackage[english]{babel} 
\usepackage{graphicx} 
\usepackage{listings} 
\usepackage{amssymb} 
\usepackage{amsmath} 
\usepackage{longtable}
\usepackage{multirow} 
\usepackage{booktabs}
\usepackage{pst-all} 
\usepackage{pstricks-add}
\usepackage{units} 
\usepackage{listings} 
\usepackage{color}
\usepackage{subfigure}
\usepackage{float}
\begin{document} 

\title{Effect of short range order on transport in one-particle, tight-binding models}

\author{Abdellah Khodja}

\email{akhodja@uos.de}

\affiliation{Fachbereich Physik, Universit\"at Osnabr\"uck,
             Barbarastrasse 7, D-49069 Osnabr\"uck, Germany}

\affiliation{Fachbereich Physik, Universit\"at Osnabr\"uck,
             Barbarastrasse 7, D-49069 Osnabr\"uck, Germany}

\author{Jochen Gemmer}

\email{jgemmer@uos.de}

\affiliation{Fachbereich Physik, Universit\"at Osnabr\"uck,
             Barbarastrasse 7, D-49069 Osnabr\"uck, Germany}

\begin{abstract}
We investigate transport properties of topologically disordered, three-dimensional,
one-particle, tight binding models, featuring site distance dependent
hopping terms. We start from entirely disordered systems into which we
gradually introduce some short range order by numerically performing a
pertinent structural relaxation using local site-pair interactions.
Transport properties of the resulting models within the delocalized regime are analyzed numerically
using linear response theory. We find that even though the generated
order is very short ranged, transport properties such as conductivity or
mean free path scale significantly with the degree of order. Mean free
paths may exceed site-pair correlation length. It is furthermore
demonstrated that, while the totally disordered model is not in accord
with a Drude- or Boltzmann-type description, moderate degrees of order
suffice to render such a picture valid.
\end{abstract}


\maketitle

\section{introduction}

Since its introduction the Anderson model has been a paradigm in the investigation of disordered quantum systems \cite{Abrahams2010}. 
However, 
most existing amorphous materials are not amorphous due to disordered on-site potentials on a periodic lattice (Anderson model)
but feature a spatially disordered site configuration. A model class for such systems has been introduced and to some extend analyzed by 
Lifshitz \cite{Lifshitz1964}. In both system classes the phenomenon of Anderson localization occurs, i.e. at some (or all) energies energy 
eigenstates extend only over a finite spatial range called the localization length. Three-dimensional systems may feature localized and 
extended 
states which are energetically separated by the mobility edges. The lowest (and highest) energy eigenstates  of an ``energy band'' 
are usually localized at all non-zero degrees of disorder, while the states in the center of the spectrum may be delocalized 
\cite{Abrahams2010}. There also 
exists a 
degree of disorder at which all eigenstates become 
localized, called the Anderson transition. While there is an enormous amount of literature on Anderson transitions   
\cite{Abrahams2010,Bauer1988}, 
mobility edges 
\cite{Knief1998, Atta-Fyn2004,Grussbach1995,priour2012,Brndiar2006} and localization lengths  
\cite{Abrahams2010,jakob2011}, there seems to be less work on the quantitative 
description of transport behavior 
(conductivities, diffusion constants, mean free paths, etc.) in the delocalized regime. This is probably due to the fact that electronic 
transport on the macroscopic scale in doped semiconductors or glassy systems is almost always dominated by thermally activated hopping 
processes between localized energy eigenstates at the lower band edge \cite{Shklovskii1984, Knief1998, Nardone2012}. At feasible 
temperatures in standard materials the fermi distribution simply gives only non-negligible probability to localized states at the lower band edge. (Highly doped but weakly compensated semiconductors may be an exception here
 \cite{Shklovskii1984, Altshuler1985}). However, transport mediated by the delocalized center of the spectrum, which is the subject of the paper at hand,  may be of importance for electronic conduction in amorphous metals or phononic heat conduction in amorphous materials
. \cite{Amir2010}. Much of the quantitative results on 
transport in the delocalized regime are either on extremely weakly disordered systems, i.e., crystalls comprising some defects 
\cite{Mertig1987, Dekker1998, Papanikolaou1994, Vojta1992} or on the Anderson model \cite{Steinigeweg2010, Markos2006, Kramer1993}. 
These investigations find localized and/or diffusive behavior in the 
limit of large time and lengthscales. remarkably diffusive and even weakly localized behavior has  been found on finite timescales 
(at high frequencies) also in strictly periodic (quantum) systems of the Lorentz gas type
. \cite{Gaspard1996, Tian2005, Tian2009}.
However,  the paper at hand addresses truely non-periodic systems and finds ballistic behavior (mean free path) on the short and diffusive 
behavior on the long lengthscale.
Recently results on transport within the delocalized regime in some Lifshitz models featuring completely random site configurations have 
been reported in Ref. 
\cite{khodja2013}.
Both transport types hopping- (though not thermally activated) and band- or Drude-transport  have been found, which provides an
alternative
to the widespread belief that 
transport phenomena within the delocalized regime in disordered systems may generally be described using a Drude or 
Boltzmann approach \cite{Altshuler1985}. The paper at
hand is along the lines of  Ref. \cite{khodja2013} and extends the studies to Lifshitz models which are not completely random but feature 
some short range order in the site configuration. We
find that even weak short range order affects transport properties strongly.

The paper at hand is organized as follows: we start by  introducing our models and their specifying parameters in Sec.
\ref{sec-re} and \ref{model}. Then we compute  in Sec. 
\ref{secconduct} the dependence of their conductivities  (at high temperatures and low fillings) on the amount of short range order.
After briefly commenting on localization and  short range order in  Sec.  \ref{localization} we address the  Einstein relation and 
mean free paths defined on the basis of an  Einstein relation in Sec. \ref{seceinstein}. By considering models featuring different length
scales of the hopping amplitudes we find some universality of the transport properties in  Sec.  \ref{transtype}.
We close with summary and conclusion in Sec. 
\ref{sumcon}.

\section{model: generation of short range order}
\label{sec-re}
\begin{figure}[h]
\centering
\includegraphics[width=7cm]{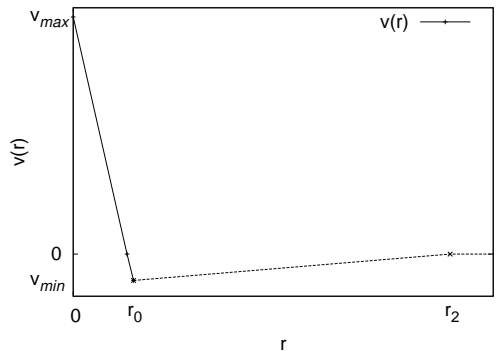}
\caption{ Polygon pair-interaction potential $v(r)$ used in the structural relaxation algorithm  (\ref{map}) to generate short range order.}
\label{fig-1}
\end{figure}

\begin{figure}[h]
\centering
\includegraphics[width=7cm]{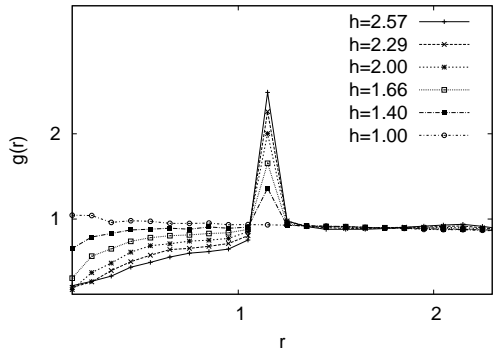}
\caption{ Pair-correlation function $g(r)$ after different run-times of the structural relaxation algorithm (\ref{map}). Obviously 
short range order is gradually generated, the most frequent site distance is stable at $r=1.12$. Based on this Figure the height of the 
first peak $h$ is used to quantify the degree of short range order. Note that the second peak is hardly visible for all degrees of order }
\label{fig-2}
\end{figure}

Even the most amorphous solids are spatially not completely random but feature some short range order on an atomic scale. As this 
order becomes 
more pronounced the amorphous system gradually passes over to a crystall. Many of those intermediate structures actually exist. It is the 
purpose of this paper to investigate the effect of increasing order in initially completely (nonphysically) disordered systems on transport 
properties. The systems will be modeled by quantum tight-binding models featuring inter-site distance-dependent hopping
amplitudes, cf. Sec. \ref{model}. Thus the site configuration eventually affects the transport properties. Indeed, as will become clear below,
changing the topological order of the atomic sites has  a substantial effect on the transport  quantities like mean free paths
conductivities. etc.  The case of 
fully disordered sites distribution was extensively investigated in Ref. \cite{khodja2013}. Thus, in the paper at hand we generate some short range
order in the following straightforward way: We start by producing a set of $N=L^3$ three-dimensional position vectors $\vec{r}_j$
by drawing each Cartesian-coordinate $(x_{j},y_{j},z_{j} )$ of each vector
independently from a uniform distribution on the interval $[0,L]$ , i.e., within a cube of
volume $L^3$ in real space. This guarantees a
uniform site distribution with unit density. Now short range order is produced based on  pair-interaction 
potentials $v(|\vec{r}_{ij}|)$ where $|\vec{r}_{ij}|=|\vec{r}_{i}-\vec{r}_{j}|$ denote interatomic distances between sites $i$ and $j$. 
We schematically mimic the relaxation which would occur through the minimization of the 
total interaction energy $V:=\sum_{ij}v(|\vec{r}_{ij}|)$ with respect to the site postions $\vec{r}_{j}$ for particles  in viscous fluid.
Routinely one could use a structural relaxation algorithm with a typical interatomic
potential such as Morse, etc.. But, due to the curvature of such potentials, the most frequent site distance grows while order is 
numerically generated. Since we intend  to exclusively focus on  the effect of the degree of order,  we want to keep other
parameters such as density, most frequent site distance,  etc fixed. Thus  we employ a rather  simple pair interaction 
potential which is essentially a polygon, see Fig. \ref{fig-1}. The parameters $r_0=1.12 ,r_2=8,v_{min}=-20,v_{max}=140$ 
control the short range repulsion and the long (intermediate) range attraction. The choice $r_0=1.12$
guarantees that throughout the lattice relaxation the value of the site density $\rho=1$  remains unchanged even if close-packing 
would be reached (which of course practically never happens). This kind of simple polygon potential may not be very realistic but it 
suffices to continuously generate a first peak in the pair-site correlation function at $r_0$, cf. Fig. \ref{fig-1}. Thus we define
our  lattice relaxation by 
the following gradient descent method:
\begin{equation}
\label{map}
x^{n+1}_{\bf i} =\left. x^{n}_{\bf i}-\lambda\frac{\partial V}{\partial x_{\bf i}}\right|_{\left\{ \displaystyle x^{n}_{\bf i} \right\}}
\end{equation}
here ${\bf i},{\bf j}$ label the Cartesian  components of all position vectors, i.e.,  ${\bf i},{\bf j}=1,....,3N$    and  n denotes 
the step-number of the minimization algorithm. 
The parameter $\lambda$ has to be adequately defined such that the algorithm is stable. This kind of algorithm will of course
not lead to a global minimum of the potential, it will rather
move the atomic sites such that the potential energy  is locally minimized. Up to a certain limit a desired degree of short range order 
may now simply be generated by iterating (\ref{map}) for a pertinent number of steps. Fig. \ref{fig-2} illustrates  the 
corresponding generated 
short range order by displaying the 
pair correlation 
function $g(r)$:
\begin{equation}
\label{cor}
g(r)=\frac{1}{4\pi r^2\rho dr}\sum_{i j} \text{rect}\left(\frac{|\vec{r}_{ij}|-r}{dr}\right)
\end{equation}
(here rect$(\cdots)$ denotes the standard rectangular function)

For small $dr$ the quantity $\sum_{i j} \text{rect}\left((|\vec{r}_{ij}|-r)/dr\right)$ should be proportional to $dr$ 
thus, the correlation function $g(r)$  is independent of the specific choice of $dr$. Unfortunately statistical effects also become 
more pronounced for smaller $dr$ since our sample is finite. Thus calculating $g(r)$ with sufficient precession may require large samples. 
We found, however, that satisfactory results may be produced from samples comprising  no more than $24^3$ sites. 
\\ 
Defining a quantity that sensibly parametrizes the degree of order in general  is a formidable task of it own. Here we exclusively focus 
on the dependence of the transport properties on the ``peak-height'' of the pair correlation function, i.e., $h=$max$(g(r))$ (which 
occurs due to our specific potential $v(r)$ always at $r=1.12$).  This peak-height assumes the value $h=1$ for the completely 
disordered system which has been addressed in detail in Ref. \cite{khodja2013} and  in principle increases to infinity for a long 
range ordered 
crystal. In this sense it may be viewed as a simple indicator for the degree of topological order in a system.

\section{MODEL: TIGHT-BINDING HAMILTONIAN}
\label{model}
Based on the short range ordered site structure described in the previous section we now specify the Hamiltonian of the model. The latter is a 
one-particle, tight-binding Hamiltonian:
\begin{equation}
\label{ham}
\hat{H}=\sum_{j k} H_{jk} \hat{a}^{\dagger}_j\hat{a}_k  
\end{equation}
$ \hat{a}_i^{\dagger},\hat{a}_i$ denotes the annihilation and creation operators.
The function $H_{jk}$ describes the dependence of the overlap or hopping amplitudes  on the positions
of the respective sites.  We  consider isotropic overlap thus $H_{jk}$ essentially
depends on the distance $ s_{jk}$   between site $j$ and site $k$. Here we specifically choose $H_{jk}$
to be a Gaussian:
\begin{equation}
 H_{jk}:= \exp\left(\frac{-4 s_{jk}^2}{\pi\tilde{l}^2}\right)
\end{equation}
The Gaussian decrease is not intended to  be specifically address any real amorphous material. It is rather motivated by numerical
feasibility: Since the system is disordered there are localized states at the edges of the energy spectrum. Those tend to become 
fewer with increasing $\tilde{l}$. For technical reasons we intend to focus on models with a negligible amount of localized states, cf. Sec. \ref{localization}.  However, for reliable results on transport from exact diagonalization on systems featuring large  
$\tilde{l}$ large 
sample sizes are needed. In Ref. \cite{khodja2013} similar systems (but featuring no short range order)  have been investigated. 
There it has been 
found that for Gaussian $H_{jk}$ a range of  $\tilde{l}$  may be found for which localization as well well as finite-size effects (at $L=24$)
are both negligible. Such a range  of  $\tilde{l}$ does not exist, e.g., for exponentially decreasing hopping-amplitudes as considered e.g., 
in Refs. \cite{priour2012, jakob2011, Bauer1988}. 
$\tilde{l}$ parametrizes the mean overlap length.  In the completely disordered system , i.e., for random sites we have:
\begin{equation}
\label{meanhop}
 \frac{1}{N}\sum_{jk} s_{jk} |H^{}_{jk}|=\tilde{l}
\end{equation}
The distances $s_{jk}$ are, due to the usage of periodic boundary conditions, somewhat complex functions. They may be defined as 
\begin{equation}
s_{jk}:=\sqrt{d^2_{jk}(x)+d^2_{jk}(y)+d^2_{jk}(z)} 
\end{equation}
where the $d$'s are essentially the Cartesian components of $(\vec{r}_j- \vec{r}_k)$. To account for periodic boundary conditions
they are specifically  defined as  
\begin{equation} 
d_{jk}(\alpha)=\begin{cases}
 |\alpha_j-\alpha_k|,  & |\alpha_j-\alpha_k|<\frac{L}{2}\\
 L-|\alpha_j-\alpha_k|,  &|\alpha_j-\alpha_k|>\frac{L}{2}
\end{cases}
\end{equation}
where $\alpha$ is one of the Cartesian coordinates, i.e., $\alpha=x,y,z$. Thus the
distance $s_{jk}$ is essentially the shortest distance between the sites $j,k$ under
periodic closure of the sample.

\section{CURRENT DYNAMICS AND CONDUCTIVITY}
\label{secconduct}
\begin{figure}[h]
\centering
\includegraphics[width=8cm]{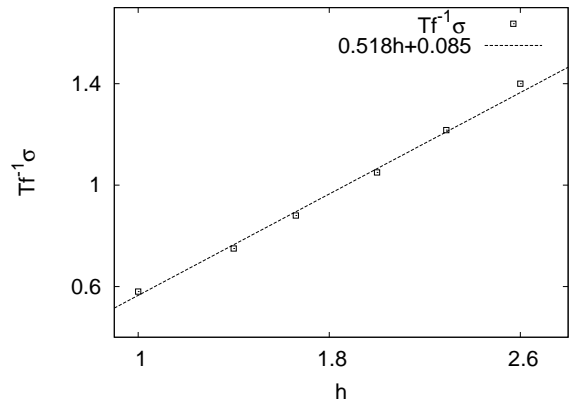}
\caption{ Scaled conductivity $T f^{-1}\sigma_{dc}$  (or diffusion constant $D$, see
 Sec. \ref{seceinstein}) for  mean overlap length $\tilde{l}=1.3$ as function of the the degree of order $h$ starting from the 
 fully disordered model $h=1$. The conductivity appears to scale linearly w.r.t. $h$,
 the dashed line is the corresponding  fit}
\label{fig-4}
\end{figure}

Now we investigate the dependence of the conductivity on $h$, i.e., different degrees of short range order.
We employ linear response theory, i.e., the  Kubo formula. In the limit of
high temperatures and low fillings (routinely described within the framework of the
grand canonical ensemble) the dc-conductivity is given as:
\begin{equation}
\label{kubo}
\sigma_{dc}= \sigma(t\rightarrow \infty), \quad   \sigma(t) =
\frac{f}{T}\int_0^{t}\frac{1}{N}\text{Tr}\{ \hat{J}(t')\hat{J}(0)\} dt'   
\end{equation}
  \cite{Kubo1991, Jaeckle1978}, where  $f$ is the filling factor (mean number of
particles  per site at equilibrium), trace and
current operators refer to the one-particle sector only, furthermore $\hat{J}(t)$ denotes
the current operator in the Heisenberg picture. $T$ is the temperature and we set
$k_B=1$, $\hbar=1$, furthermore we set the charges of the particles to unity, i.e.,
$q=1$. Now of course an appropriate current operator has to be defined. In the
context of periodic systems and next neighbor hoppings this is often done by considerations based on the
continuity equation for the particle density  \cite{Zotos1999, Heidrich-Meisner2003,
Benz2005, Gemmer2006}. Here we choose a  definition of the  the current which is
based on the ``velocity'' in, say, $x$-direction. Eventually this choice will be justified by the agreement of the results 
with the diffusion constant in the sense of a Einstein relation, cf.  Sec. \ref{seceinstein}. The velocity operator reads:
\begin{equation}
\label{vel}
\hat{v}=\text{i}[\hat{H},\hat{x}]
\end{equation}
Here $\hat{x}$ is a $x$-position operator and it is defined as 
\begin{equation}
\label{pos}
\hat{x}= \sum_{i=1}^{N}x_i\hat{n}_i, \quad  \hat{n}_i:= \hat{a}_i^{\dagger}\hat{a}_i
\end{equation}
where $x_i$ is the $x$-coordinate of the position of site $i$. Thus, the operator
$\hat{v}$ may also be written as 
\begin{equation}
\label{velocity}
\hat{v}=  \text{i}\sum_{ij}(x_j-x_i)H_{ij}  \hat{a}_i^{\dagger}  \hat{a}_j     
\end{equation}
The interpretation of such an operator as velocity or current is not in entire agreement with periodic boundary conditions.
A (slow) transition of probability from, say, the right edge of the sample ($x=L$)
to the left edge of the sample ($x=0$) would give rise to very high negative
velocities. But within the concept of periodic boundary conditions such a transition
should correspond to low positive velocities (across the boundary). Thus in order to obtain a suitable
current operator we modify the above velocity operator (\ref{velocity}) such that it
features the same structure for transitions arising from the periodic closure as it
already exhibits for transitions within the sample:
\begin{equation}
 \hat{J}=\sum\limits_{ij} J_{ij} \hat{a}^\dagger_i \hat{a}_j
\end{equation}
\begin{figure}[h]
\centering
\includegraphics[width=8cm]{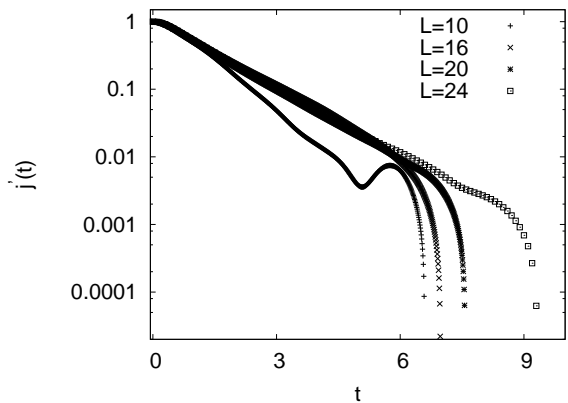}
\caption{ Normalized current auto-correlation function $j'(t)$ for mean overlap lenght $\tilde{l}=1.3$ and short range order quantified by 
$h=2.57$ for increasing sample sizes $L$. Since the graphs coincide in regions where they are substantially different from zero, for, say 
$L \geq 16$, data can reliably be expected to contain negligible finite-size effects at $L=24$. Moreover,  the linear dependence 
of the current 
auto-correlation function on time in the logarithmic plot suggest an exponential decay which indicates 
Boltzmann transport  }
\label{fig-3}
\end{figure}
\begin{equation*}
 J_{ij}=\begin{cases}
 \text{i}[x_j-x_i]H_{ij}  &|x_j-x_i|<\frac{L}{2} \\
 \mathrm{sgn}(x_j-x_i)\left[i[L-|x_j-x_i|]H_{ij}\right]  &|x_j-x_i|>\frac{L}{2} 
\end{cases}
\end{equation*}
Equipped with this definition for the current we may now simply calculate the
current auto-correlation function as appearing in (\ref{kubo}). We do so using
standard numerically exact diagonalization routines. Within reasonable computing
time we are able to treat samples up to a size of $L=24$. In order to be able to compare the key features of the dynamics of  the
current auto-correlation functions for various degrees of order and model sizes to each other
we compute a kind of ``normalized'' current auto-correlation function,
$j'(t):=\text{Tr}\{ \hat{J}(t)\hat{J}(0)\}/\text{Tr}\{ \hat{J}^2(0)\}$. Before analyzing conductivity and transport behavior, 
we briefly address finite-size effects and numerical limitations. We find that for all models discussed in the paper at hand sample 
sizes of  $L=24$ are sufficient to get rid of significant finite-size effects. This is illustrated exemplarily in  Fig. \ref{fig-3}.
The  normalized current correlation functions $j'(t)$ for the different sample sizes above, say, $L=16$ 
coincide for the relevant initial times, at which $j'(t)$ is substantially  different from zero, hence the finite-size 
 effects are indeed negligible. For $L=24$ a matrix of dimension $d\approx 14 000$ has to be diagonalized and a corresponding correlation 
function has to be computed. This  is  numerically feasible but demanding on standard computers.
 In order to analyze  conductivity we plot the ``scaled'' conductivity $\sigma_{dc} T/f$ for various generated short range orders at fixed mean 
 overlap length $\tilde{l}=1.3$ against $h$, see Fig. \ref{fig-4}. 
The plot clearly suggest a linear dependence of the conductivity on the peak height of the site pair-correlation function. The 
corresponding fits yield for the respective conductivities:

\begin{equation}
\label{conduc}
\sigma_{dc}(\tilde{l}=1.3)=\frac{f}{T}\left ( 0.518h+0.085\right ). 
\end{equation}
This equation implies that for increasing short range order the conductivity increases significantly. If the most frequent site distance is 
only twice more frequent that any other long range distance, the conductivity is roughly doubled compared to  the completely random model.
This means that even in the regime of amorphous systems a slight increase of order will affect transport properties substantially. 
Furthermore considerations based on Fig. \ref{fig-5} may 
indicate a transition from a ``Non-Drude'' to Boltzmann- or Drude- type of transport. If one computes a current-correlation function 
from a Drude model or a Boltzmann equation (in relaxation-time approximation) one always obtains an exponential decay of the current. Thus,
in order for some (quantum) dynamics to be in accord with Drude-type of model, it must yield an exponentially decaying 
current-correlation function. In the model at hand, however, exponentially decaying  current-correlation functions only appear at a certain 
degree of short range order. To illustrate this we plot the normalized current-correlation function $j'(t)$ in Fig. \ref{fig-5} for $h=1$ 
(complete disorder) and $h=2.57$\\
At $h=1$ the curve agrees well with a Gaussian fit. Such a decay of the current cannot result from a Boltzmann equation. 
The latter may  yield multi-exponential decay if behavior beyond the relaxation time approximation is taken into account, but no Gaussian 
relaxation.
However, at the short range order specified by  $h=2.57$ the decay gradually  
passes over to an  exponential as illustrated by the respective 
exponential fit. This implies a transition from ``Non-Drude'' to Drude transport.
\begin{figure}[h]
\centering
\includegraphics[width=7.5cm]{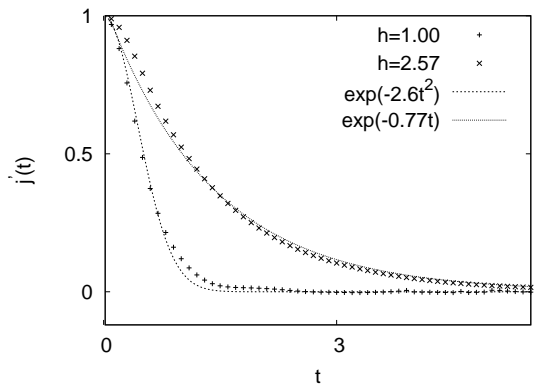}
\caption{Normalized current correlation function $j'(t)$ at  mean overlap length $\tilde{l}=1.3$ as function of 
time t for the fully disordered model $h=1$ and the short range ordered model  $h=2.57$. At $h=1$ the decay appears to be approximately 
Gaussian whereas decay and at  $h=2.57$ it is dominantly exponential (cf. Fig.  \ref{fig-3}) which indicates Boltzmann transport.} 
\label{fig-5}
\end{figure}
 Note that this transition occurs still in the 
strongly disordered regime, even at $h=2.57$ the system its far away from a crystal containing some impurities. The existence of this 
transition may be supported by an investigation of the dependence of a mean free path on the short range order. 
Such an investigation is presented in  Sec. \ref{seceinstein}.

\section{Comment on localization}
\label{localization}

\begin{figure}[h]
 \subfigure[]{\includegraphics[width=7.5cm]{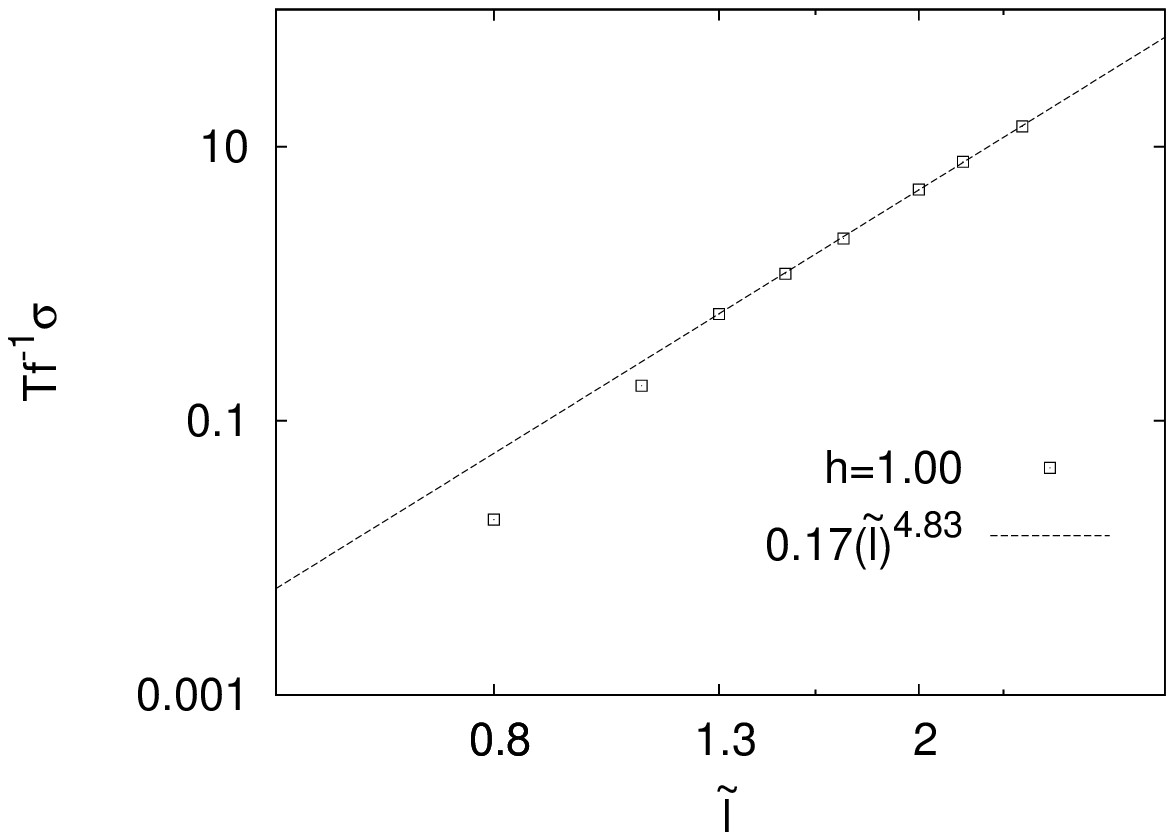}
\label{fig-6a}
}
  \subfigure[]{\includegraphics[width=7.5cm]{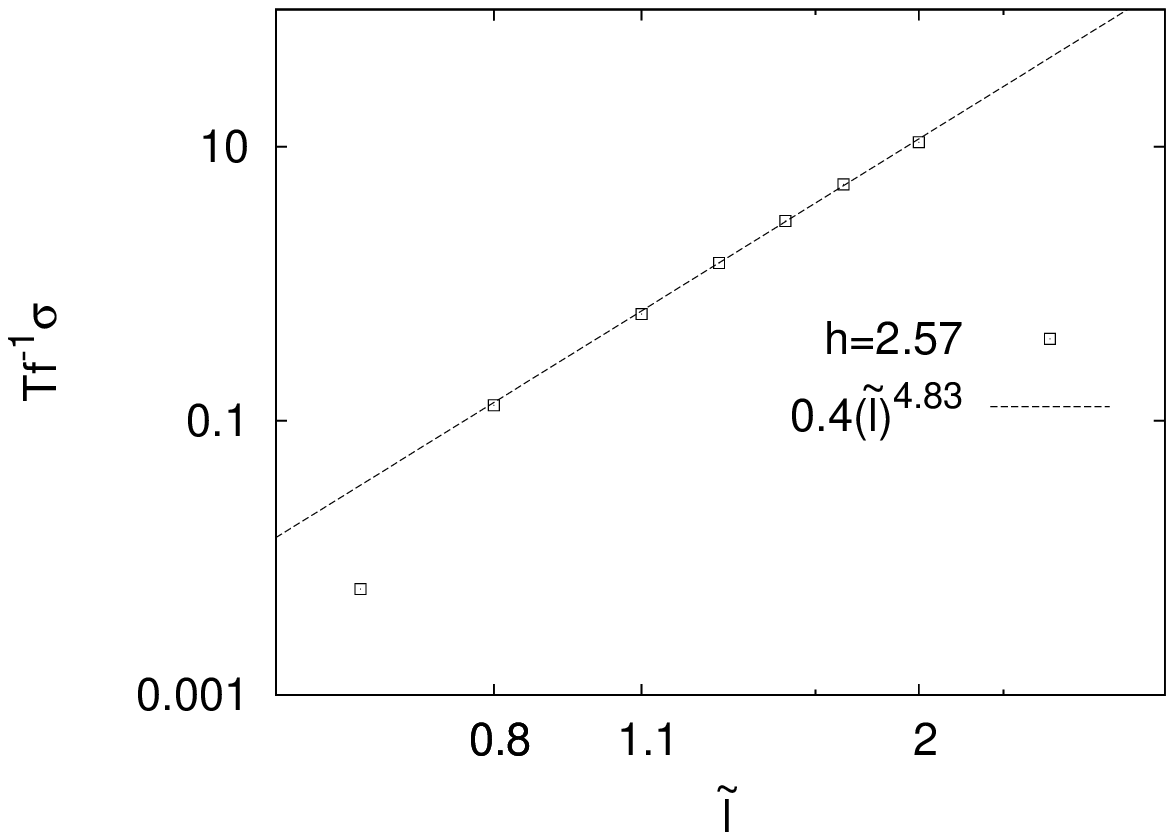}
\label{fig-6b}
}
\caption{ (double logarithmic plot) Panel (a) shows  the scaled conductivity $T f^{-1}\sigma_{dc}$ for the fully disordered model $h=1$. Already at overlap lengths 
as long as $\tilde{l}=1.1$ 
there are deviations from the power law. This indicates localization of substantial parts of the spectrum.
Panel (b) shows the scaled conductivity $T f^{-1}\sigma_{dc}$ at  $h=2.57$. The power law is fulfilled down to  the overlap 
length $\tilde{l}=0.8$ which indicates that almost all states are delocalized at an overlap length as short as $\tilde{l}=0.8$}
\label{fig-6}
\end{figure}

In Ref. \cite{khodja2013} it was found, using methods based on the inverse participation ratio, that in the topological fully 
disordered model ($h=1$)
at  mean overlap length $\tilde{l} \geq 1.3$ almost the entire spectrum is delocalized. 
For smaller overlap lengths more and more energy eigenstates become localized. The Anderson transition,  at which the entire
spectrum is localized, occurs  roughly at $\tilde{l}\approx 0.6$. The same work furthermore reports that the conductivity scales as a 
power law with mean overlap lengths in the fully delocalized regime, i.e., for ($\tilde{l} \geq 1.3$):
\begin{equation}
\label{pow}
\sigma_{dc}(h=1.00)=\frac{f}{T}0.17\tilde{l}^{4.83}
\end{equation}
In the paper at hand we computed the conductivity for even smaller  mean overlap lengths, $\tilde{l}<1.3$, see Fig. \ref{fig-6}.

 Obviously,  deviations from the power law appear right below  $\tilde{l}\approx 1.3$, i.e., at the point at which substantial parts of the 
 spectrum become 
 localized. Those deviations increase rapidly for decreasing mean overlap length. 
 Thus, we interpret the deviations from the power law (\ref{pow}) as a consequence of increasing localization. This is supported by
 investigations based on inverse
participation ratio  \cite{khodja2013,priour2012}.
 We now use those findings  to produce a rough estimate for the localization properties of
 the various short range ordered models. To those ends we compute the conductivities for different mean overlap lengths for $h=2.57$ and use 
 the deviation 
 from the power law as an indicator for the onset of substantial localization. Indeed Fig. \ref{fig-6b} shows that for the short range 
 ordered model the conductivity satisfies the power law (\ref{pow}) down to $\tilde{l} \approx 0.8$. Below that deviations from the power 
 law arise. Thus we conclude that the onset of substantial localization occurs in this short range ordered model at an even lower  mean
 overlap length, namely  $\tilde{l} \approx 0.8$. This fits into the overall picture since one expects in the limit of fully ordered systems 
 (crystals) delocalization to occur for arbitrarily small overlap length $\tilde{l}$. This finding suggest that probably also in the 
 respective localized regime localization lengths are longer in the presence of short range order. A conclusive statement on this as well as 
 on the universality class of the short range ordered models is, however, beyond the scope of the paper at hand and thus left for 
 further research.

\section{Einstein relation and mean free paths}
\label{seceinstein}
Apart from the conductivity the diffusion coefficient is another important transport
quantity. According to the Einstein relation conductivity and diffusion constant
should be proportional to each other. However, the validity of the Einstein relation
and the limits of its applicability have been  much debated subjects and continue to
be so in the context of quantum systems \cite{Steinigeweg2009} (and references
therein). Recently it has been reported that the Einstein relation holds for
periodic, interacting, 1-d quantum systems at high temperatures. It is claimed to
hold even for finite times, thus taking the form 
\begin{equation}
\label{einsteinepl}
D(t)=\frac{T}{\epsilon^2} \sigma(t)
\end{equation}
here $D(t)$ is the (time-dependent) diffusion constant, $\epsilon^2$ is the
uncertainty (variance) of the transported quantity per site at the respective
equilibrium \cite{Steinigeweg2009}. In Ref. \cite{khodja2013} it has been demonstrated that  (\ref{einsteinepl}) also holds for completely disordered systems of 
the type 
considered in the paper at hand. Furthermore an analytical argument for the validity of (\ref{einsteinepl}) has been presented which does not 
depend on the topological structure at all. However, since this argument is not conclusive  we investigate in the following numerically 
whether
 (\ref{einsteinepl}) also holds for
short range ordered systems. In our case the
transported quantity is the particle density. In the limit of high temperatures and
low fillings the equilibrium fluctuations scale as $\epsilon^2=f$  \cite{Kubo1991}.
Thus if one hypothetically accepts the validity of (\ref{einsteinepl}) also for the
systems at at hand, one gets from inserting (\ref{kubo}): 

\begin{figure}[h]
\centering
\includegraphics[width=7.5cm]{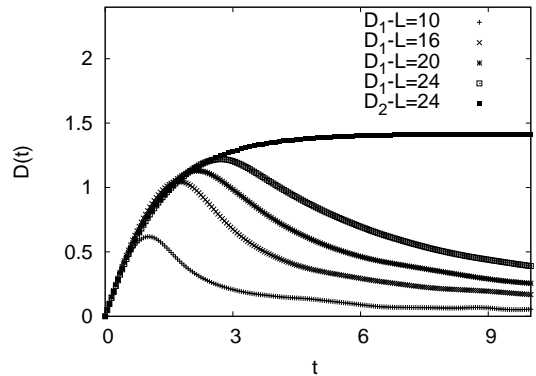}
\caption{Comparison of two methods to calculate (time-dependent) diffusion
coefficients: $D_1(t)$ from (\ref{diffone}) and   $D_2(t)$ from (\ref{difftwo}). The
data address the generated short range order quantified by $h=2.57$  and  mean overlap 
length $\tilde{l}=1.3$. Obviously, finite-size effects are more pronounced for $D_1(t)$; however, it appears to coincide with 
$D_2(t)$ up to increasing times for increasing sample sizes. This coincidence implies the validity of an
Einstein relation.} 
\label{fig-7}
\end{figure}

\begin{equation}
\label{einsteinone}
D(t)=\int_0^{t}\frac{1}{N}\text{Tr}\{ \hat{J}(t')\hat{J}(0)\} dt'
\end{equation}
If a diffusion
equation holds, the derivative w.r.t. time of the spatial variance of the diffusing
quantity equals twice the diffusion constant  \cite{Steinigeweg2009}. We numerically analyze
the dynamics of this variance using an initial state of the form 
\begin{equation}
\label{init}
 \rho(0) =\frac{1}{Z}\exp{(-\frac{(\hat{x}-\frac{L}{2})^2}{2})}, \quad 
Z= \text{Tr}\{ \exp{(-\frac{(\hat{x}-\frac{L}{2})^2}{2})}    \}
\end{equation}
i.e., a state in which the probability is more or less concentrated in a thin slab
of a thickness on the order of one, perpendicular to the $x$-axis in the middle of
the cubic sample. We calculate the increase of the variance of this state and take a
derivative w.r.t. time, thus obtaining directly a diffusion constant which we call $D_1(t)$:
\begin{equation}
\label{diffone}
D_1(t)=\frac{1}{2}\frac{d}{dt}\text{Tr}\{ \hat{x}^2(t) \rho(0) \}
\end{equation}
(Note: the particle density does not drift hence $\frac{d}{dt}\text{Tr}\{ \hat{x}(t) \rho(0) \}=0$)
We compare this to the r.h.s. of  (\ref{einsteinone}) which should equal the diffusion coefficient if the Einstein relation holds, 
thus we call this quantity $D_2(t)$: 
\begin{equation}
\label{difftwo}
D_2(t)=\int_0^{t}\frac{1}{N}\text{Tr}\{ \hat{J}(t')\hat{J}(0)\} dt' 
\end{equation}
If the  Einstein relation holds $D_1(t)$ and  $D_2(t)$ should coincide.

The results are displayed in Fig. \ref{fig-7}. Although finite-size effects are much more pronounced for $D_1(t)$ than for  $D_2(t)$
there is a  good agreement during an initial time period. This period obviously increases with system size. More specifically, 
Fig. \ref{fig-7} suggest that the time during which  $D_1(t)$ and  $D_2(t)$ coincide becomes arbitrarily long for arbitrarily large systems
Thus we conclude that the Einstein relation is valid for coherent one-particle transport in the short range ordered systems at hand.

\begin{figure}[h]
\centering
\includegraphics[width=7.5cm]{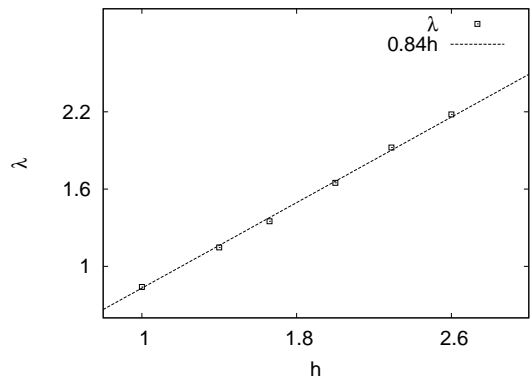}
\caption{Mean free path $\lambda$ as function of the degree of short range order $h$ at the mean overlap lenght $\tilde{l}=1.3$. Obviously the mean free path increases
significantly with increasing short range order. } 
\label{fig-8}
\end{figure}

The coincidence of $D_1(t)$ and  $D_2(t)$ allows for a definition of a mean free path $\lambda$ on the basis of  $D_2(t)$ which is, as 
demonstrated above computationally less demanding. The mean free path is introduced as follows: If the particle was completely ballistic 
(infinite mean free path)
the current auto-correlation function would never decay and the time-dependent
diffusion coefficients in the sense of (\ref{einsteinone}) would always increase
linearly. The time-dependent diffusion 
coefficients  of the models at hand increase linearly at the beginning, cf. Fig.
\ref{fig-7}, but reach a final plateau after that initial period. We
define, somewhat arbitrarily, the ballistic period as the period before the
diffusion coefficient has reached $90\%$ of its eventual value. Now we call the mean
free path the square root of the increase of the spatial variance of an initial
state of type (\ref{init}) during this ballistic period. So the mean free path is
roughly the initial increase of  width of an initially  narrow probability
distribution up to the point where the fully diffusive dynamics begins. In this way a mean free path may be defined even in the Non-Drude 
regime where traditional notions of mean free paths do not apply
 \cite{Kubo1991}. However in the Drude regime this definition roughly coincides with
traditional mean free path. The so defined mean free paths $\lambda$ are displayed in Fig. \ref{fig-8} for short range ordered models featuring different $h$ 
(but fixed $\tilde{l}=1.3)$. The mean free
path appears  to increase linearly  with $h$. Although the generated topological order is small and the 
structure is still near a fully disorder model the mean free path increases 
substantially with respect to  $h$. In the Drude regime this may be viewed as corresponding to a decrease of the scattering cross section.
 This leads to the remarkable situation that the mean free path exceeds the range of the order, e.g., 
at $h=2.57$, recall that the most frequent site distance has been kept fixed at $r_0=1.12$ and a second peak is hardly visible in 
Fig. \ref{fig-2}. Thus,  we conclude that ballistic motion of particles is not necessarily restricted to the range of order as often 
assumed. These findings suggest that transport behavior for  these short range ordered models may be described by a Drude model or 
a Boltzmann equation for, say, $h>2.6$ as already indicated in Sec. \ref{secconduct}. This Drude-transport is then much alike  
the dynamics of a particle in a periodic lattice featuring some impurities or a system of quasi free, weakly interacting particles. 
At $h=1$, however, the mean free path is below the most frequent site distance $r_0=1.12$. This Non-Drude transport is comparable  
to the dynamics of an over-damped
Brownian particle or the  thermally activated hopping transport which may occur in the localized regime of amorphous or /and doped 
semiconductor \cite{Shklovskii1984}. Again, this is in accord with the findings in  Sec. \ref{secconduct}. It may be worth pointing out that 
both transport types have also
been found in other one-particle quantum systems, e.g., Non-Boltzmann transport  in modular
quantum systems  \cite{Weaver2006, Michel2005} and both transport-types in the three-dimensional 
Anderson model  \cite{Steinigeweg2010, Brndiar2006}. Note that while any dynamics featuring a finite mean free path yield 
diffusive behavior described by some conductivity like displayed in
 Fig. \ref{fig-4} on the macroscopic scale, the concrete size of the 
mean free path will alter transport through structures that are on the order of the mean free path significantly. thus transport 
through thin films or nanostructures may quantitatively depend on the mean free path.

\section{Transport behavior for varying overlap lengths}
\label{transtype}

\begin{figure}[h]
\centering
\includegraphics[width=7.5cm]{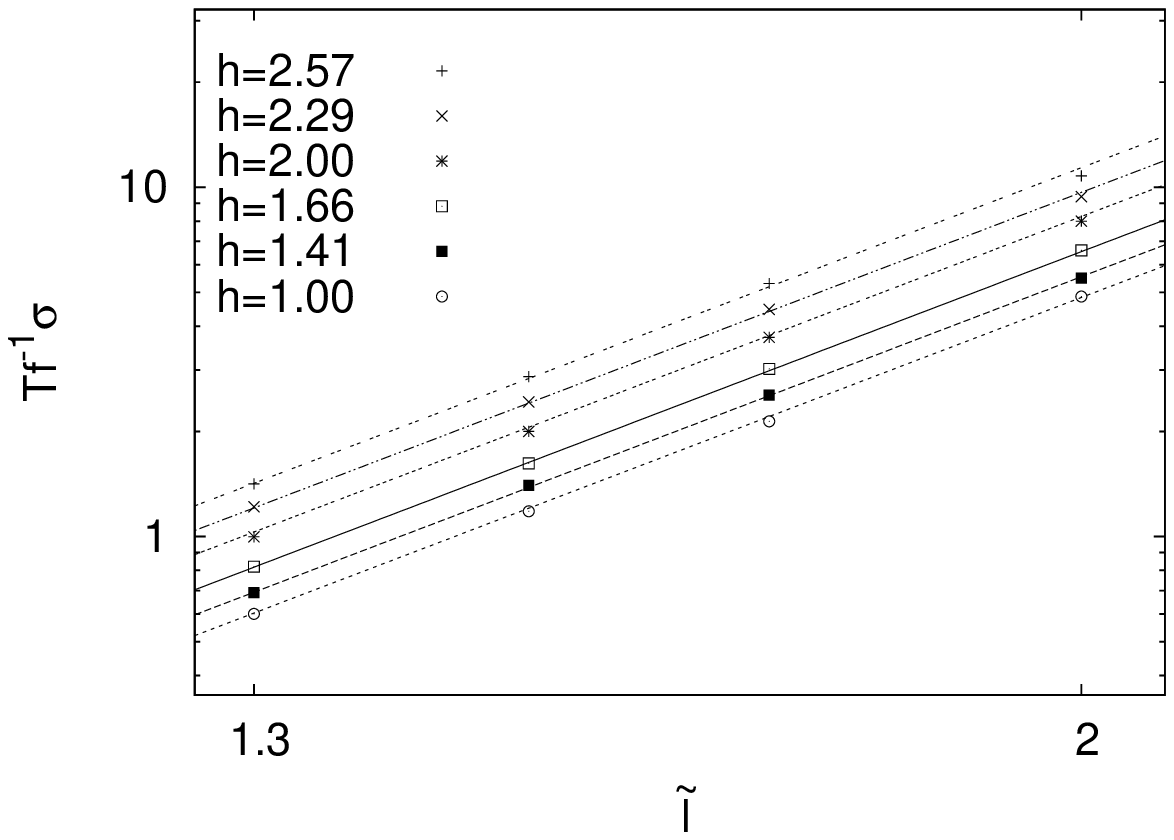}
\caption{ Scaled conductivities $T f^{-1}\sigma_{dc}$  as functions of mean overlap length $\tilde{l}$ 
for various degrees of  short range order parametrized by $h$
on a double logarithmic scale. For all $h$ the conductivities appear to follow the same 
power law w.r.t. $\tilde{l}$; the dashed ($---$) lines are the corresponding fits. This points in the direction of an universality. } 
\label{fig-9}
\end{figure}

\begin{figure}[h]
\centering
\includegraphics[width=7.5cm]{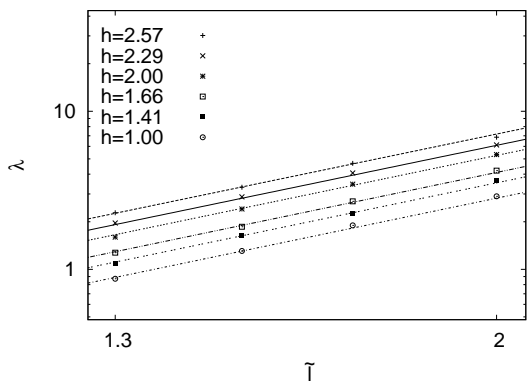}
\caption{Mean free paths $\lambda$  as functions of mean overlap length $\tilde{l}$ 
for various degrees of  short range order parametrized by $h$
on a double logarithmic scale. For all $h$ the mean free paths appear to follow the same 
power law w.r.t. $\tilde{l}$; the dashed ($---$) lines are the corresponding fits. This points in the direction of an universality. } 
\label{fig-10}
\end{figure}

Until now we studied solely the effect of increasing short range order at fixed mean overlap length $\tilde{l}=1.3$. The latter is the 
shortest   $\tilde{l}$ at which almost all energy eigenstates are delocalized, even for the completely disordered model \cite{khodja2013}.
Our method is not suitable to investigate even shorter  $\tilde{l}$ since it does not resolve w.r.t energy (high temperature limit). 
The investigation
of larger   $\tilde{l}$ is, however, to some extend possible. Thus in this Section we investigate the dependence of transport parameters on
both the amount of order, $h$, and the mean overlap length, $\tilde{l}$. We use the same method as described in the 
previous  Sections, eg., linear response theory. The results are displayed in Figs. \ref{fig-9} and  \ref{fig-10}.

Obviously, Fig. \ref{fig-9} exhibits a power-law dependence of the conductivity on $\tilde{l}$, with the same exponent for all 
$h$. More specifically Fig. \ref{fig-10} suggests the following form of the conductivity within the investigated range of $h,\tilde{l}$:

\begin{equation}
\label{cod}
\sigma_{dc}(\tilde{l},h)=\frac{f}{T}\left ( 0.146h+0.024\right ) \, \, \tilde{l}^{4.83}, \quad  
\end{equation}
This product form indicates a kind of universality: whatever the amount of short range order is, the scaling with the mean overlap length is 
always the same and vice versa. A similar situation is found for the scaling of the mean free path $\lambda$. Fig. \ref{fig-10} suggests:
\begin{equation}
\label{free1}
\lambda (\tilde{l},h)=(0.42 h) \tilde{l}^{2.68}, \quad  
\end{equation}

Whether or not this universality holds for even more different model types is a tentative subject of 
further research.

\section{Summary and conclusion}
\label{sumcon}
We investigated the transport behavior of a class of quantum systems which may be described as
three-dimensional, topologically short range ordered, one-particle, tight-binding
models. These models are meant to be very
simplified descriptions of amorphous materials in the delocalized  regime. 
Conductivity and mean free paths at low fillings and high temperatures have been determined essentially by evaluating
the Kubo formula using numeric solutions of the Schroedinger equation for finite samples comprising up to $\approx 14 000$ sites.
Conductivities and mean free paths are  found to scale linearly with a measure of the (low) amount of order  
and as a power law with the mean overlap length of the hopping amplitudes. The fact that conductivity and mean free path appear to be 
product functions
w.r.t. to those parameters indicates a kind of universality. The scaling with order is such that mean free paths which exceed the range of
order are reached at comparatively low degrees of order. This is interpreted as a transition towards a Boltzmann or Drude type of transport,
i.e., almost free, weakly scattered particles, in a rather amorphous regime. We furthermore verified the validity  of an 
Einstein relation for those systems and found explicit hints that increasing order pushes the Anderson transition towards shorter
mean overlap lengths. The latter findings are in accord with generic expectations. 

\section{acknowledgements}

We thank H. Niemeyer for fruitful discussions.


%

\end{document}